\documentclass[amsmath,amssymb,amsbsy,prb,twocolumn,preprintnumbers,showpacs,superscriptaddress]{revtex4-2}
\usepackage{graphicx,color}
\usepackage{dcolumn}
\usepackage{bm}
\usepackage{braket}
\usepackage{mathtools}
\usepackage{ulem}
\usepackage[breaklinks,colorlinks=true,linkcolor=blue,urlcolor=blue,citecolor=blue]{hyperref}

\begin{document}
\title{Flat-band solutions in $D$-dimensional decorated diamond and pyrochlore lattices: Reduction to molecular problem}
\author{Tomonari Mizoguchi}
\affiliation{Department of Physics, University of Tsukuba, Tsukuba, Ibaraki 305-8571, Japan}
\email{mizoguchi@rhodia.ph.tsukuba.ac.jp}
\author{Hosho Katsura}
\affiliation{Department of Physics, University of Tokyo, Hongo, Bunkyo-ku, Tokyo 113-0033, Japan}
\affiliation{Institute for Physics of Intelligence, The University of Tokyo, 7-3-1 Hongo, Bunkyo-ku, Tokyo 113-0033, Japan}
\affiliation{Trans-scale Quantum Science Institute, University of Tokyo, Bunkyo-ku, Tokyo 113-0033, Japan}
\author{Isao Maruyama}
\affiliation{Department of Information and Systems Engineering, Fukuoka Institute of Technology, Fukuoka 811-0295, Japan}
\author{Yasuhiro Hatsugai}
\affiliation{Department of Physics, University of Tsukuba, Tsukuba, Ibaraki 305-8571, Japan}

\begin{abstract}
Flat-band models have been of particular interest from both fundamental aspects 
and realization in materials.
Beyond the canonical examples such as 
Lieb lattices and line graphs, a variety of tight-binding models are found to possess flat bands.
However, the analytical treatment of dispersion relations is limited, 
especially when there are multiple flat bands with different energies.
In this paper, we present how to determine flat-band energies and wave functions
in tight-binding models on decorated diamond and pyrochlore lattices in generic dimensions $D\geq 2$.
For two and three dimensions, such lattice structures are relevant to various organic and inorganic materials, 
and thus our method will be useful to analyze the band structures of these materials. 
\end{abstract}
\maketitle

\section{Introduction \label{sec:intro}}
Singular dispersions in band structures
are the source of a variety of interesting phenomena in solid-state physics.
One of the representative examples is a linear dispersion around the band crossing point, or the Dirac/Weyl point~\cite{Vafek2014,Armitage2018,Bernevig2018},
which gives rise to various intriguing transport~\cite{Novoselov2006,Fukushima2008,Zyuzin2012} and magnetic~\cite{Fukuyama1970,Koshino2007,Raoux2015,Maebashi2017} phenomena.
As such, Dirac/Weyl fermions in solids have been intensively 
pursued~\cite{vonNeumann1929,Hatsugai2010,Asano2011,Vafek2014,Armitage2018,Bernevig2018}.
Another example of singular dispersion is a flat band,
which is a completely dispersionless band in the entire Brillouin zone.
Studies of such band structure have been developed in various aspects,
such as ferromagnetism~\cite{Mielke1991,Tasaki1992,Mielke1993,Kusakabe1994,Tasaki1998,Tamura2019,Tasaki2020}, 
superconductivity~\cite{Imada2000,Kuroki2005,Kobayashi2016,Matsumoto2018,Aoki2019},
topological phenomena~\cite{Aoki1996,Vidal1998,Guo2009,Weeks2010,Katsura2010,Green2010,Tang2011,Sun2011,Neupert2011,Sheng2011,Wang2011,Liu2012,Pal2018,Rhim2019,Mizoguchi2020,Kuno2020,Kuno2020_2},
and localization phenomena~\cite{Goda2006,Chalker2010,Bilitewski2018,Kuno2020_3,Danieli2020,Orito2020}.

So far, various tight-binding models with flat bands 
have been explored~\cite{Mielke1991,Sutherland1986,Miyahara2005,Hatsugai2011, Maimaiti2017,Misumi2017,Lee_Fleurence2019,Maimaiti2019,Mizoguchi2019, Mizoguchi2019_2,Maimaiti2021}, and
many insights on the model construction have been accumulated.
It was also found that some flat-band models have large sublattice degrees of freedom, 
resulting in multiple flat bands with different energies~\cite{Hatsugai2015,Fujii2018,Fujii2019,Mizoguchi2019_3}. 
In such models, it is not easy to obtain analytic expressions of 
dispersion relations since the Hamiltonians in momentum space are large matrices.

\begin{figure}[b]
\begin{center}
\includegraphics[clip,width = 0.95\linewidth]{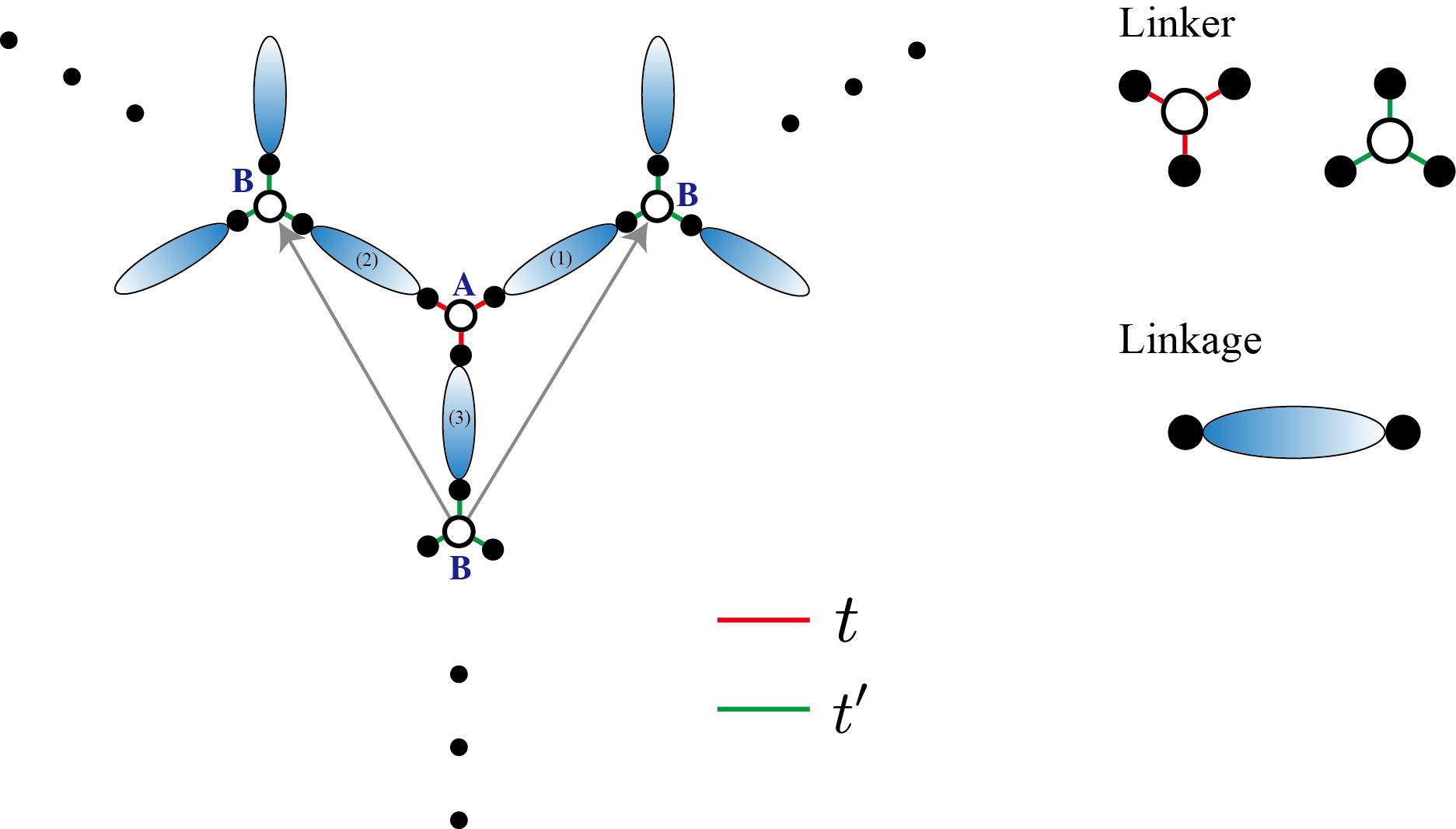}
\vspace{-10pt}
\caption{A schematic of generic decorated diamond lattices.
For clarity, we draw the case with $D=2$.
White dots stand for the vertices of the original diamond lattice, 
and gray allows stand for the lattice vectors.
Blue ellipses and black dots denote decorated parts.
The schematics of linkers and linkages are also depicted. Note that the black dots 
belong to both linkers and linkages. }
  \label{fig:schematic}
 \end{center}
 \vspace{-10pt}
\end{figure}

In this paper, we elucidate how to determine the flat-band energies analytically 
in a class of tight-binding models 
which can be obtained by decorating the bonds of 
a honeycomb lattice (in two dimensions), and a diamond lattice (in three dimensions),
and their higher-dimensional analogs, $D \geq 4$; see Fig.~\ref{fig:schematic} for the schematic figure of the two-dimensional model. 
Such lattice structures are of interest because
they are known to be realized in various organic-based materials, 
such as graphene superstructures~\cite{Shima1993,Morishita2021},
$\alpha$-graphyne~\cite{Baughman1987,Longuinhos2014,Li2015,Barreteau2017},
and metal-organic frameworks (MOFs)~\cite{Liu2013,Yamada2016,Barreteau2017,Kumar2018},
as well as some inorganic materials~\cite{Lee2019,Park2019}.
Recently, they were also discussed in the context of the square-root topological phases~\cite{Mizoguchi2020_2,Mizoguchi2020_3}.
We therefore expect that the determination of the flat-band energies is useful for band structure analysis and material design for these materials. 

The key idea is to divide the Hamiltonian into two parts,
which we term ``linkers" and ``linkages".
Importantly, the linkers and the linkages are not independent of each other since they share sites.
Nevertheless, the flat-band energies can be obtained by solving the linkage Hamiltonian, 
and the corresponding wave function can be found 
such that the compatibility relations are respected on the shared sites.  
The momentum-independence of the eigeneneries originate from the fact that 
the linkage Hamiltonian can be regarded as that for an isolated ``molecule"~\cite{Katsura2015}. 
We find that the flat-band wave function of the $D$-dimensional decorated diamond lattice 
is given by the product of the wave function of the linkage and 
the flat-band wave function of $D$-dimensional pyrochlore lattice.
We also shed light on another interesting band structure often seen in this class of lattices,
namely, a multiple band touching at $\Gamma$ point which occurs at specific choices of parameters.

The analog of the method described in this paper was previously applied 
to two-dimensional decorated kagome lattice~\cite{Mizoguchi2019_3}, 
relevant to covalent organic frameworks (COFs)~\cite{Fujii2018} as well as the cyclicgraphdiyne~\cite{You2019},
where carbon atoms having different kinds of $sp$ hybrid orbitals coexist and form a crystal.
Here we emphasize that this method yields not only 
the energies of the flat bands but also their wave functions,
which was not addressed in the previous work.  
Therefore, for completeness, we also explain the method for 
obtaining flat-band energies and eigenstates for  
the $D$-dimensional decorated pyrochlore lattice.

In the following discussions in the main text, we impose several assumptions 
(see Sec.~\ref{sec:formulation_M}) in order to retain the relevance to real materials.
However, from a theoretical point of view, some of the assumptions can be relaxed.
Such generalizations are described in Sec.~\ref{sec:summary} as well as Appendix~\ref{app:extensions}.

\section{Flat-band solutions for $D$-dimensional decorated diamond lattices \label{sec:formulation}}
In this section, we first describe the decorated diamond model, which is the main focus of this paper. 
We then explain how the flat-band energies and wave functions can be determined.
The key idea is to employ 
a technique of mathematical physics by which we can reduce the eigenvalue problem of the Bloch Hamiltonian with a relatively large size
to that of the small molecule.
\begin{figure}[tb]
\begin{center}
\includegraphics[clip,width = 0.95\linewidth]{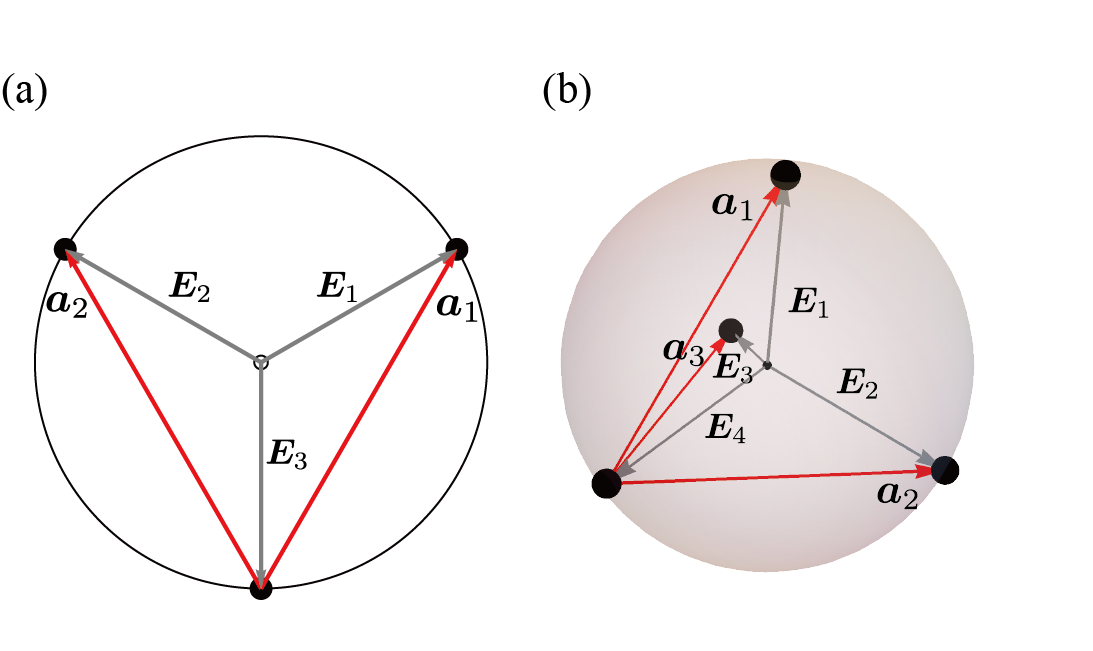}
\vspace{-10pt}
\caption{Schematics of $\bm{a}_j$ and $\bm{E}_j$ of Eq.~(\ref{eq:def_aande}) for (a) two and (b) three dimensions.}
  \label{fig:def_vector}
 \end{center}
 \vspace{-10pt}
\end{figure}
\subsection{Model \label{sec:formulation_M}}
Consider a diamond lattice in $D$ dimensions with $D \geq 2$~\cite{Hatsugai2011,Creutz2008,Kimura2010,Kato2017}.
The lattice vectors are given as~\cite{Hatsugai2011}
\begin{eqnarray}
\bm{a}_j =\bm{E}_j- \bm{E}_{D+1}, \label{eq:def_aande}
\end{eqnarray}
where $j=1,\cdots,D$ and the vectors $\bm{E}_{1}, \cdots, \bm{E}_{D+1}$ 
are the vertices of the $D$-simplex;
see Fig.~\ref{fig:def_vector} for the schematics of $D=2,3$.
We set the coordinates of two sublattices of the $D$-dimensional diamond lattice
(for $D=2$, see the white dots of Fig.~\ref{fig:schematic}) as
\begin{eqnarray}
\bm{r}_{\rm A}  = \frac{1}{D+1} \sum_{j=1}^{D} \bm{a}_j, 
\end{eqnarray}
and 
\begin{eqnarray}
\bm{r}_{\rm B} =\bm{0}.
\end{eqnarray}

Now, let us consider the decorated lattices of $D$-dimensional diamonds, 
shown in Fig.~\ref{fig:schematic}. 
Namely, we decorate the nearest-neighbor (NN) bonds of the diamond lattices,
obeying the following rules:
\begin{itemize}
\item The sublattices A and B of the original diamond lattice are, respectively, connected to 
$D+1$ sites with the same hoppings (red and green bonds in Fig.~\ref{fig:schematic}).
\item The decorated objects are the same for all the NN bonds of the diamond lattices
[see Eq.~(\ref{eq:linkage_same}) for details]. 
\end{itemize}

For later use, let us clarify some terminologies:
\begin{itemize}
\item We call a set of $D+2$ sites, composed of one site placed on the original diamond lattice 
and the other $D+1$ sites connected to that site, a ``linker".
\item We call a decorated part on each edge of the diamond lattice a ``linkage". 
In other words, the linkages are placed on the vertices of the line graph of the diamond lattice.
\end{itemize}
It is worth noting that the black dots in Fig.~\ref{fig:schematic} belong to both a linker and a linkage.
  
On this class of lattices, we consider the following Hamiltonian in $\bm{k}$-space,
which is in general written as a $[(D+1)q + 2]$-dimensional matrix:
\begin{widetext}
\begin{eqnarray}
\mathcal{H}_{\bm{k}} = 
\begin{pmatrix}
U_{\rm A} & t \bm{x}_q^{\rm T} &  t \bm{x}_q^{\rm T} & \cdots &  t \bm{x}_q^{\rm T}& 0 \\
 t \bm{x}_q &\mathcal{H}^{(1)}_{\mathrm{linkage}} & \mathcal{O}_q & \cdots & \mathcal{O}_q & t^\prime e^{i\bm{k}\cdot \bm{a}_1} \bm{y}_q\\
t \bm{x}_q  & \mathcal{O}_q &  \mathcal{H}^{(2)}_{\mathrm{linkage}} & \ddots &\vdots & t^\prime e^{i\bm{k}\cdot \bm{a}_2} \bm{y}_q \\ 
  \vdots    &   \vdots & \ddots  & \ddots  & \mathcal{O}_q & \vdots \\
  t \bm{x}_q & \mathcal{O}_q & \cdots & \mathcal{O}_q&  \mathcal{H}^{(D+1)}_{\mathrm{linkage}} & t^\prime \bm{y}_q \\
 0& t^\prime e^{-i\bm{k}\cdot \bm{a}_1} \bm{y}_q^{\rm T} & t^\prime e^{-i\bm{k}\cdot \bm{a}_2} \bm{y}_q^{\rm T} & \cdots &  t^\prime  \bm{y}_q^{\rm T} & U_{\rm B} \\
\end{pmatrix}, \label{eq:ham_gen}
\end{eqnarray}
\end{widetext}
where $\bm{x}_q$ and $\bm{y}_q$ are $q$-component column vectors 
defined as 
$\bm{x}_q:= (1,0,\cdots ,0)^{\rm T}$ and $\bm{y}_q:= (0,\cdots ,1)^{\rm T}$, respectively,
and $\mathcal{O}_q$ stands for the $q\times q$ zero matrix.
The parameters, $U_{\rm A}$ and $U_{\rm B}$, are on-site potentials for sublattices A and B, respectively;
$t$ and $t^{\prime}$ are, respectively, 
the transfer integrals assigned on the bonds connecting the decorated part with the sublattices A and B.

For later use, we define two $(D+1)$-component row vectors, 
$\bm{\psi}^{(1) \dagger}$ and 
$\bm{\psi}^{(2) \dagger}_{\bm{k}}$,
which have the forms:
\begin{subequations}
\begin{eqnarray}
\bm{\psi}^{(1) \dagger}= \left(1,1,\cdots,1 \right),
\end{eqnarray}
and
\begin{eqnarray}
\bm{\psi}^{(2) \dagger}_{\bm{k}} = \left(e^{-i\bm{k} \cdot \bm{a}_1}, \cdots, e^{-i\bm{k}\cdot \bm{a}_{D}},1 \right),
\end{eqnarray}
\end{subequations}
and the matrix composed of these two vectors~\cite{remark}:
\begin{eqnarray}
\Psi_{\bm{k}}^\dagger = 
\begin{pmatrix}
\bm{\psi}^{(1) \dagger} \\
\bm{\psi}^{(2) \dagger}_{\bm{k}} \\
\end{pmatrix}.
\end{eqnarray}

The $q \times q$ matrix $\mathcal{H}^{(j)}_{\mathrm{linkage}}$ in Eq.~(\ref{eq:ham_gen})
can be regarded as a Hamiltonian of an isolated ``molecule".
In the present case, we assume that all the linkages have the same structure, i.e., the following holds:
\begin{eqnarray}
\mathcal{H}^{(1)}_{\mathrm{linkage}} = \mathcal{H}^{(2)}_{\mathrm{linkage}}  = \cdots = \mathcal{H}^{(D+1)}_{\mathrm{linkage }}
= \mathcal{H}_{\mathrm{linkage}}. \label{eq:linkage_same}
\end{eqnarray}

\subsection{Derivation of flat-band solution}
This type of models possess multiple flat bands with different energies~\cite{Shima1993,Barreteau2017,Lee2019}.
Remarkably, if the number of decorated sites on each bond is $q$, 
there exist $q$ flat bands with different energies.
More precisely, in the $D$-dimensional model, each flat band has $(D-1)$-fold degeneracy,
thus the number of flat bands is equal to $(D-1)q$. 

In general, analytic solutions of the dispersion relations in this class of models 
are hard to obtain, 
since the size of the Hamiltonian matrix is large. 
Nevertheless, we can obtain the eigenvalues and eigenvectors of the flat bands as follows.
Let $\bm{\lambda}_{\mathrm{linker},\bm{k}}$ be a $(D+1)$-component column vector, which satisfies
\begin{eqnarray}
\Psi_{\bm{k}}^\dagger \bm{\lambda}_{\mathrm{linker},\bm{k}} = 
\begin{pmatrix}
0 \\
0\\
\end{pmatrix}.
\label{eq:def_lambda}
\end{eqnarray}
As $\Psi_{\bm{k}}^\dagger$ is the $2 \times (D+1)$ matrix, 
there are $D-1$ independent solutions of $\bm{\lambda}_{\mathrm{linker},\bm{k}}$.
Only at the $\Gamma$ point (i.e., $\bm{k}=\bm{0}$), the rank of $\Psi_{\bm{k}}^\dagger$ is reduced by one as 
$\bm{\psi}^{(1)} = \bm{\psi}^{(2)}_{\bm{k}}$ holds,
which results in the increase of the number of solutions from $D-1$ to $D$. 
We note that $\bm{\lambda}_{\mathrm{linker},\bm{k}}$ 
corresponds to the flat-band eigenvector of the $D$-dimensional pyrochlore lattice~\cite{Hatsugai2011}.

To find the flat-band solution, we employ a notion of ``intertwiner"~\cite{DiRrancesco1990,Pearce1993}.
Before going to the concrete problem, we briefly address a generic argument.
Let $A$ and $G$ be Hermitian matrices with different sizes. 
It is known that $A$ and $G$ have common eigenvalues 
if these matrices satisfy
\begin{eqnarray}
A C = C G,
\end{eqnarray}
with $C$ being a non-square matrix.
The matrix $C$ is called the ``intertwiner".
A simple proof of this statement is as follows. 
Let $\bm{\phi}$ be an eigenvector of $G$ with eigenvalue $\varepsilon$. 
Then, one finds that $C \bm{\phi}$ is an eigenvector of $A$ with eigenvalue $\varepsilon$ 
(unless $\bm{\phi}$ belongs to the kernel of $C$),
because
\begin{eqnarray}
A (C \bm{\phi} ) = CG \bm{\phi} = \varepsilon (C \bm{\phi} ). 
\end{eqnarray}

Turning to the present model, we can explicitly construct the intertwiner 
$C_{\bm{k}}$ which is $[(D+1)q  +2] \times q $ matrix and satisfies
\begin{eqnarray}
\mathcal{H}_{\bm{k}} C_{\bm{k}} = C_{\bm{k}} \mathcal{H}_{\mathrm{linkage}}.  \label{eq:intertwiner_fb}
\end{eqnarray}
Its form is given as
\begin{eqnarray}
C_{\bm{k}} = 
\begin{pmatrix}
 \bm{0}_{q}^{\rm T}  \\
[\bm{\lambda}_{\mathrm{linker},\bm{k}}]_{1} I_{q}  \\
\vdots \\
[\bm{\lambda}_{\mathrm{linker},\bm{k}}]_{D+1} I_{q}  \\
 \bm{0}_{q}^{\rm T}  \\
\end{pmatrix}, 
\label{eq:intertwiner}
\end{eqnarray}
where $\bm{0}_{q}$ stands for the $q$-component column zero vector, $I_q$ stands for the $q\times q$ identity matrix,
and $[\bm{\lambda}_{\mathrm{linker},\bm{k}}]_{j}$ is the $j$-th component of $\bm{\lambda}_{\mathrm{linker},\bm{k}}$. 
Therefore, the eigenvalues of $\mathcal{H}_{\mathrm{linkage}}$ are also those of $\mathcal{H}_{\bm{k}}$. 
As $\mathcal{H}_{\mathrm{linkage}}$ is $\bm{k}$-independent, the eigenvalues obtained as such 
naturally form flat bands.
Equation~(\ref{eq:intertwiner_fb}) also leads to the flat-band wave function.
Let $\bm{\phi}_{\mathrm{linkage},n}$ be a $q$-component vector 
which is the $n$th eigenvector of $\mathcal{H}_{\mathrm{linkage}}$.
It satisfies
\begin{eqnarray}
 \mathcal{H}_{\mathrm{linkage}} \bm{\phi}_{\mathrm{linkage},n} 
 = \varepsilon_{\mathrm{linkage},n}\bm{\phi}_{\mathrm{linkage},n}
\end{eqnarray}
with $\varepsilon_{\mathrm{linkage},n}$ being the eigenvalue.
Then, the flat-band eigenvector $\bm{\varphi}_{\bm{k},n}$, written as
\begin{eqnarray}
\bm{\varphi}_{\bm{k},n} = \begin{pmatrix} 
\varphi_{A,\bm{k},n} \\ 
\varphi_{1,\bm{k},n} \\
\vdots \\
\varphi_{(D+1)q,\bm{k},n} \\
\varphi_{B,\bm{k},n} \\
\end{pmatrix}, 
\end{eqnarray}
is given as
\begin{eqnarray}
\bm{\varphi}_{\bm{k},n}  = \frac{1}{\mathcal{N}_{\bm{k}}}
C_{\bm{k}} \bm{\phi}_{\mathrm{linkage},n},
\end{eqnarray}
where $\mathcal{N}_{\bm{k}}$ is the normalization constant. 
More concretely, the components of $\bm{\varphi}_{\bm{k},n}$ are given as
\begin{eqnarray}
\varphi_{A,\bm{k},n}  = \varphi_{B,\bm{k},n}  = 0,
\end{eqnarray}
and 
\begin{eqnarray}
\varphi_{q (j-1) + m ,\bm{k},n} = \frac{1}{\mathcal{N}_{\bm{k}}} [\bm{\lambda}_{\mathrm{linker},\bm{k}}]_j  [\bm{\phi}_{\mathrm{linkage},n}]_m\label{eq:wave_function}
\end{eqnarray}
with $j=1,\cdots D + 1 $ and $m=1,\cdots q$. 
Equation~(\ref{eq:wave_function}) indicates that
the flat-band wave function of the $D$-dimensional decorated diamond lattice 
is given by the product of the linkage's wave function 
and the flat-band wave function of $D$-dimensional pyrochlore lattice.
 
In the next section, we elucidate how this construction actually works by showing specific examples. 

\section{Examples}
In this section, we demonstrate 
that the aforementioned method works for decorated diamond lattices in $D=2$, $3$ and $4$.
Although our formulation is applicable to generic types of decoration patterns, 
we mainly focus on the model where the chain-type structure is inserted between the neighboring sites of the diamond lattices.
(We present an example of the non-chain-type decorating sites for $D=2$; see Fig.~\ref{fig:dhsq}.)
The motivation to focus on these models is that, for $D=2,3$, 
they are known to be relevant to 
MOFs such as DCBP$_3$Co$_2$ and DCA$_3$Co$_2$~\cite{Kumar2018},
$\alpha$-graphyne~\cite{Baughman1987,Longuinhos2014,Li2015,Barreteau2017},
and TaS$_2$~\cite{Lee2019,Park2019}.
(DCBP and DCA stand for dicyanobiphenyl and dicyanoanthracene, respectively).
As for $D=4$, some recent works addressed the four-dimensional diamond lattice~\cite{Creutz2008,Kimura2010,Kato2017} 
as a canonical example of four-dimensional Dirac fermions on lattice models.
In this context, the decorated four-dimensional diamond lattice is an interesting extension of it where Dirac fermions and flat bands coexist.
 
 \subsection{Two dimensions: Decorated honeycomb lattice \label{sec:honeycomb}}
\begin{figure*}[t]
\begin{center}
\includegraphics[clip,width = 0.95\linewidth]{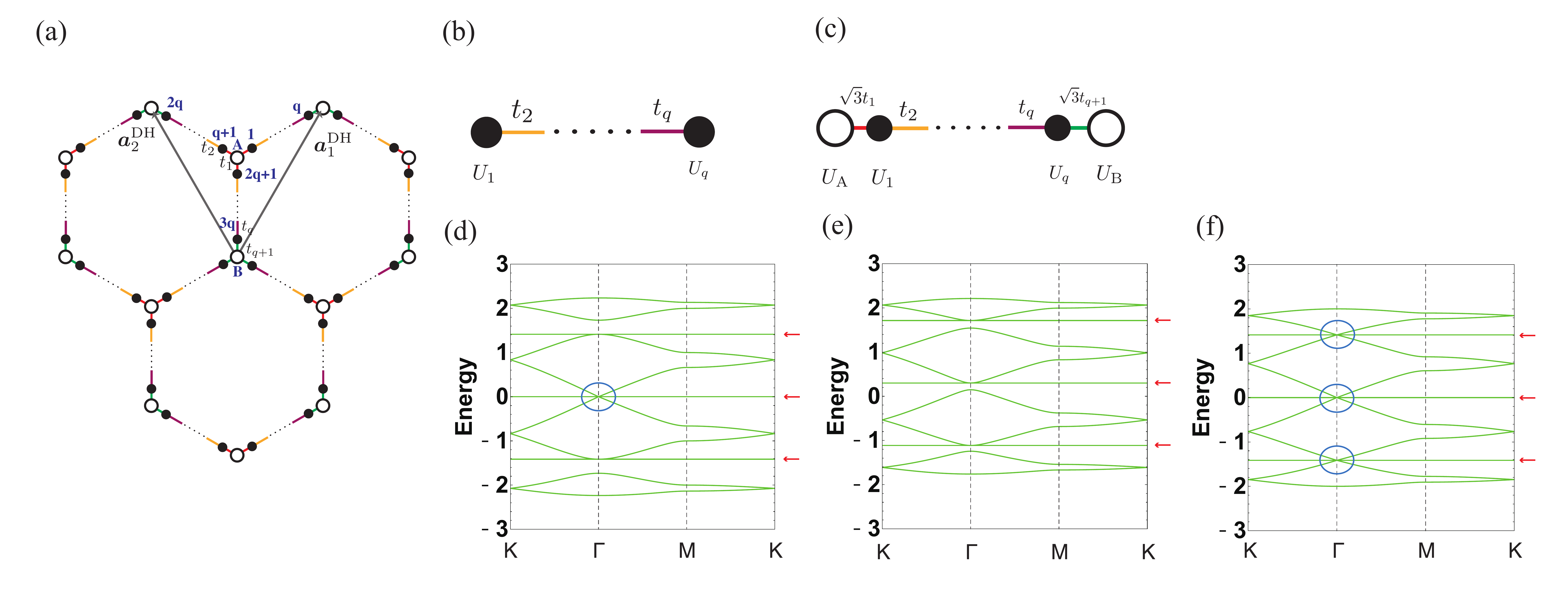}
\vspace{-10pt}
\caption{
(a) A decorated honeycomb lattice with $q$ sites on edges of hexagons.
The lattice vectors are $\bm{a}^{\rm DH}_1 = \left(\frac{1}{2}, \frac{\sqrt{3}}{2} \right)$ and $\bm{a}^{\rm DH}_2 = \left(-\frac{1}{2}, \frac{\sqrt{3}}{2} \right)$.
Schematics of the Hamiltonians of the chain-like molecules corresponding to (b) $\mathcal{H}^{\rm DH}_{\mathrm{linkage}}$
and (c) $\tilde{\mathcal{H}}$. 
The band structures for $q=3$
for (d) 
$(t_1,t_2,t_3,t_4,U_{\rm A},U_1,U_2,U_3,U_{\rm B}) = (1,1,1,1,0,0,0,0,0)$, 
(e) $(0.8,1,1,0.8,0,0.3,0.3,0.3,0)$, and (f) $\left(\sqrt{\frac{2}{3}},1,1,\sqrt{\frac{2}{3}},0,0,0,0,0 \right)$.
Red arrows point to the flat bands, and the blue circles represent the triple band touchings.
The coordinates of the high-symmetry points in the first Brillouin zone are 
$\Gamma = (0,0)$, $K = \left(\frac{4\pi}{3}, 0\right)$, and $M = \left(\pi, \frac{\pi}{\sqrt{3}} \right)$.}
  \label{fig:1}
 \end{center}
 \vspace{-10pt}
\end{figure*}
Consider a decorated honeycomb lattice model with 
$q$ sites on each edge of hexagons [Fig.~\ref{fig:1}(a)].
The specific form of the Hamiltonian, $\mathcal{H}^{\rm DH}_{\bm{k}}$, is given by substituting 
$t = t_1$, $t^\prime = t_{q+1}$, and 
\begin{eqnarray}
\mathcal{H}^{\rm DH}_{\rm linkage}
= 
\begin{pmatrix}
U_1 & t_2 & &&& \\
t_2 & U_2 & t_3 &&& \\
 & t_3 & U_3 & \ddots & \\
& &  \ddots &\ddots & \\
& & & & U_{q-1} & t_q \\
&&&& t_q & U_q \\
\end{pmatrix}
 \label{eq:ham_dh}
\end{eqnarray}
into Eq.~(\ref{eq:ham_gen}).
The row vectors $\bm{\psi}^{\mathrm{DH}(1)  \dagger}$ and $\bm{\psi}^{\mathrm{DH}(2)  \dagger}_{\bm{k}}$ are given as
\begin{eqnarray}
\bm{\psi}^{\mathrm{DH}(1)  \dagger} = (1,1,1),
\end{eqnarray}
\begin{eqnarray}
\bm{\psi}^{\mathrm{DH}(2)  \dagger}_{\bm{k}} = (e^{-i \bm{k}\cdot \bm{a}^{\rm DH}_1}, e^{-i \bm{k}\cdot \bm{a}^{\rm DH}_2},1).
\end{eqnarray}

\begin{figure}[b]
\begin{center}
\includegraphics[clip,width = 0.95\linewidth]{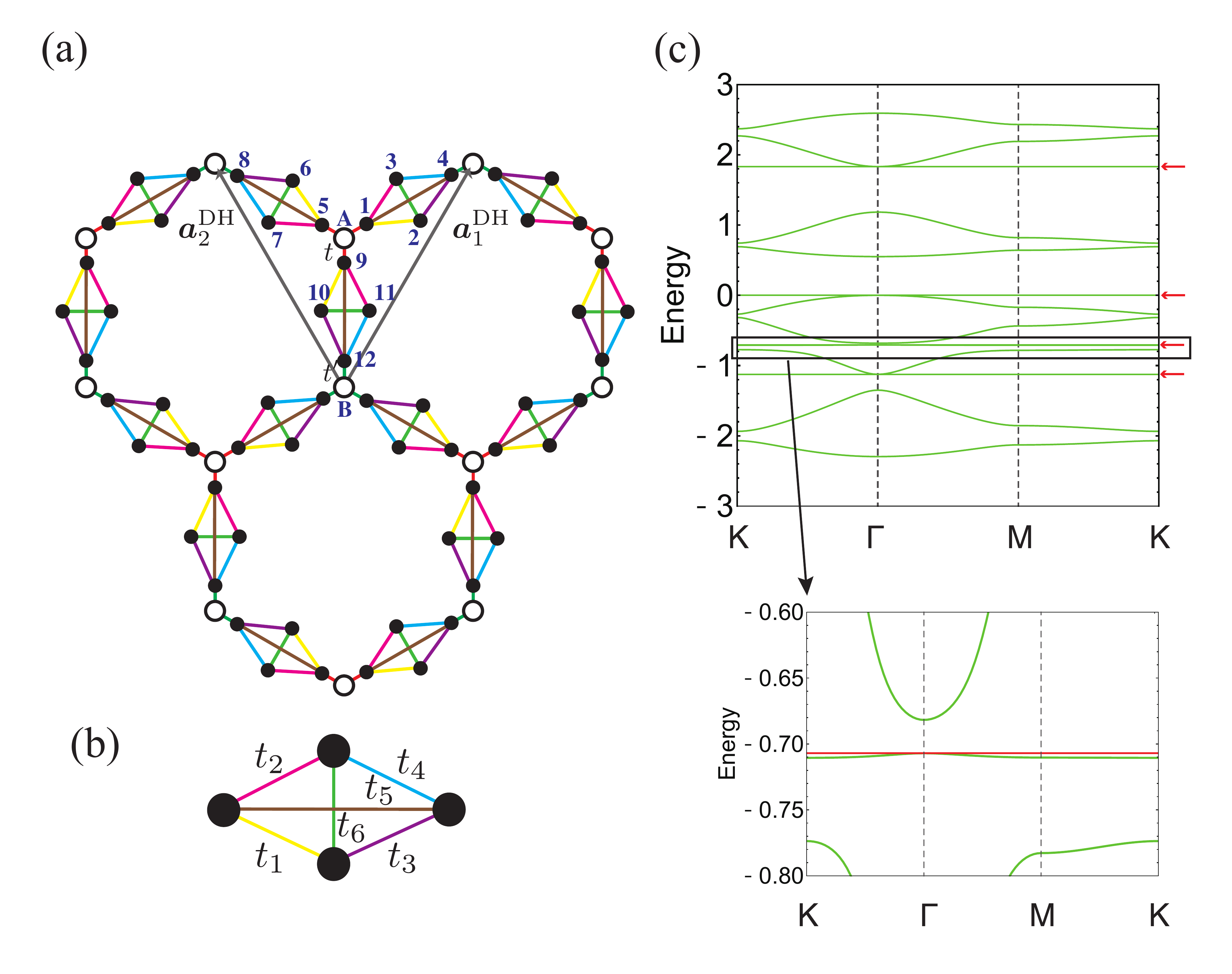}
\vspace{-10pt}
\caption{(a) A decorated honeycomb lattice with four decorating sites, shaped in the rhombus, at each edge. 
(b) Schematic figure of the four-site molecule whose eigenenergies are equal to the flat-band energies.
(c) The band structure for $(t,t^\prime, t_1,t_2,t_3,t_4,t_5,t_6) = (1, 0.9, 0.4,0.5,0.3, 0.6, 1.1, 0.7)$.
The on-site potentials are zero. 
Red arrows point to the flat bands.}
  \label{fig:dhsq}
 \end{center}
 \vspace{-10pt}
\end{figure}

The vector $\bm{\lambda}^{\rm DH}_{\mathrm{linker},\bm{k}}$ is obtained as
\begin{eqnarray}
\bm{\lambda}^{\rm DH}_{\mathrm{linker},\bm{k}} = 
\left(
\begin{array}{c}
1 - e^{-i\bm{k}\cdot \bm{a}^{\rm DH}_2} \\
e^{-i\bm{k}\cdot \bm{a}^{\rm DH}_1}-1 \\
e^{-i\bm{k}\cdot \bm{a}^{\rm DH}_2} -e^{-i\bm{k}\cdot \bm{a}^{\rm DH}_1}\\
\end{array}
\right). 
\label{eq:lambda_dh}
\end{eqnarray}
Then, the flat-band energies are equal to the eigenvalues of $\mathcal{H}^{\rm DH}_{\mathrm{linkage}}$,
and the corresponding wave functions are given in the form of Eq.~(\ref{eq:wave_function}).

In Figs.~\ref{fig:1}(d)-(f), we plot the band structures for $q= 3$ with several sets of parameters.
In all cases, there are three flat bands, whose energies are indeed equal to $\varepsilon_{\mathrm{linkage},n}$. 
\begin{figure}[b]
\begin{center}
\includegraphics[clip,width = 0.95\linewidth]{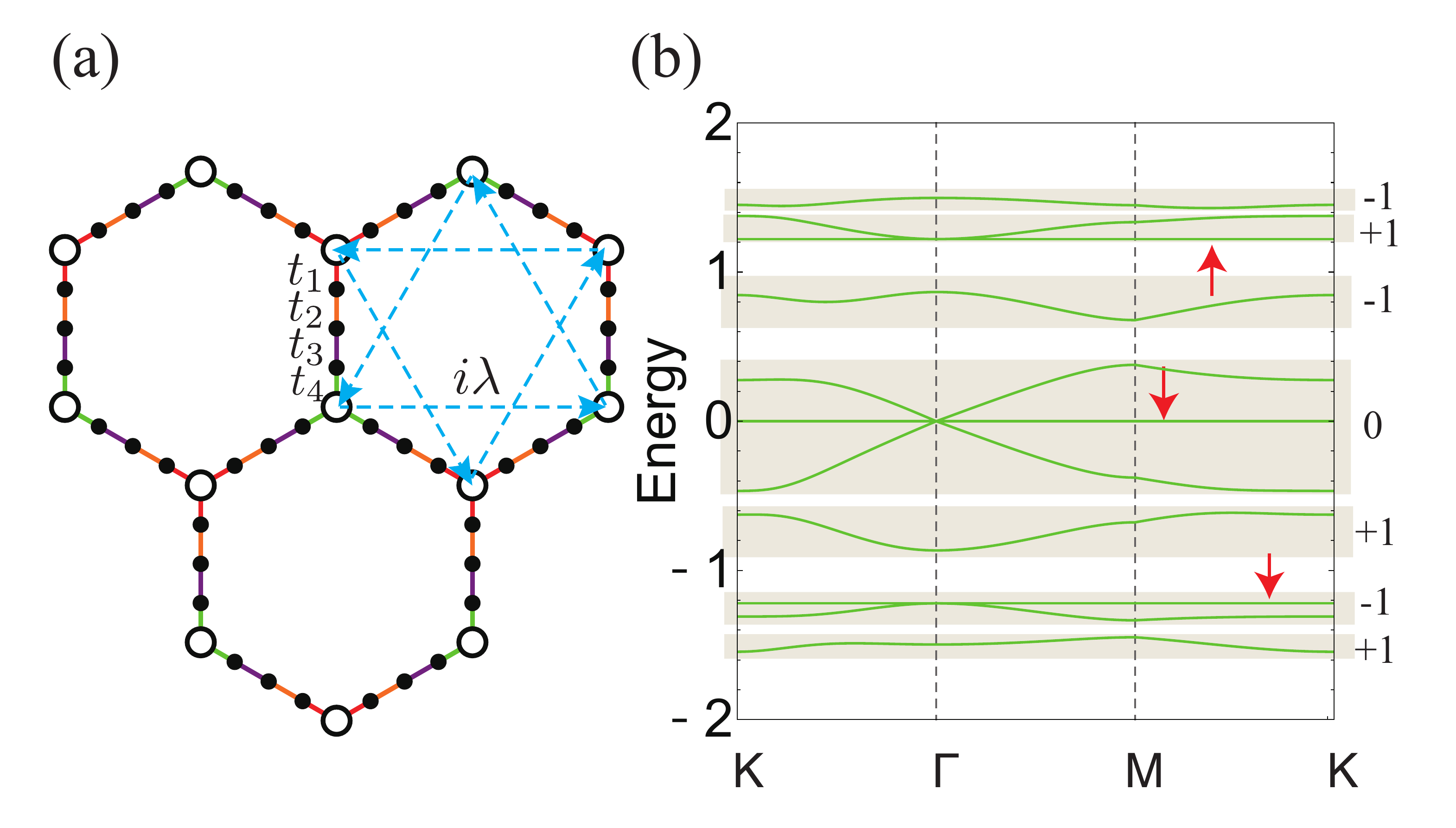}
\vspace{-10pt}
\caption{
(a) Schematic figure of $\mathcal{H}^\prime_{\bm{k}}$ of Eq.~(\ref{eq:Hp}).
Blue dashed arrows represent the complex hoppings.
(b) Band structure for the model of (a) with $(t_1,t_2,t_3,t_4,\lambda) = (0.5,0.7,1.0,0.5,0.1)$.
The on-site potentials are zero. 
The numbers beside the bands indicate the Chern numbers, which are calculated for the set of bands included in the same shade.
Red arrows point the flat bands.}
\label{fig:Chern}
\end{center}
\vspace{-10pt}
\end{figure}
\begin{figure*}[t]
\begin{center}
\includegraphics[clip,width = 0.95\linewidth]{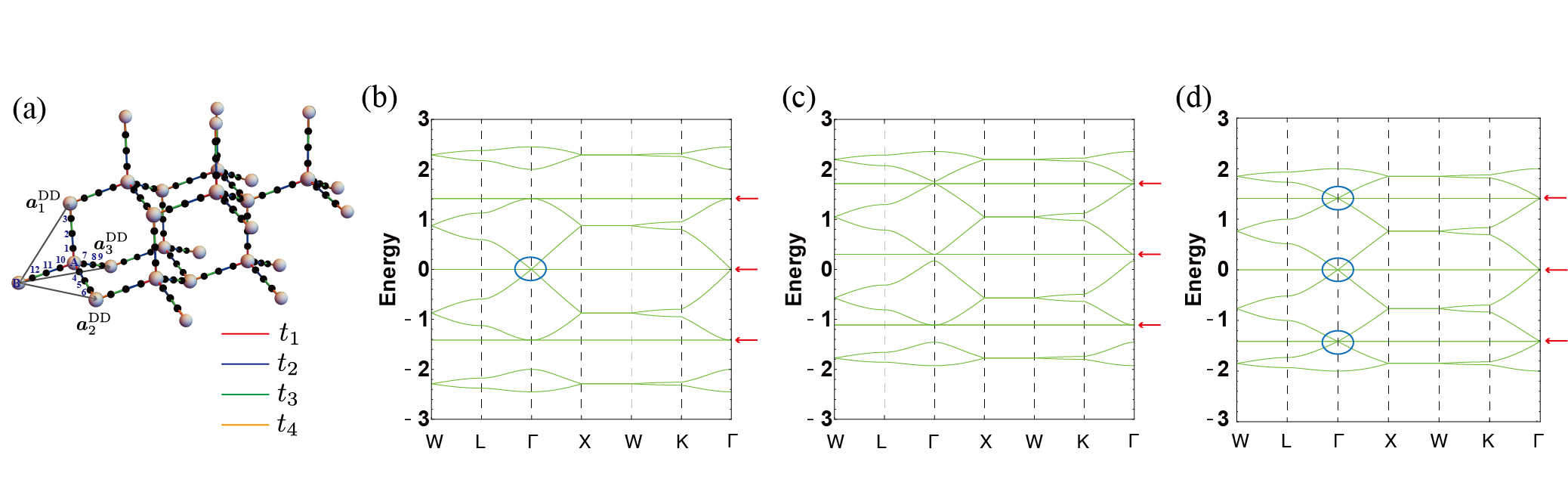}
\vspace{-10pt}
\caption{
(a) A decorated diamond lattice with three sites on NN bonds of a diamond lattice.
The lattice vectors are $\bm{a}_1^{\rm DD} = \left(0, \frac{1}{2}, \frac{1}{2} \right)$, 
$\bm{a}_2^{\rm DD} = \left(\frac{1}{2},0, \frac{1}{2} \right)$, 
and $\bm{a}_3^{\rm DD} = \left(\frac{1}{2}, \frac{1}{2},0 \right)$.
The band structures for (b) 
$(t_1,t_2,t_3,t_4,U_{\rm A},U_1,U_2,U_3,U_{\rm B}) = (1,1,1,1,0,0,0,0,0)$,
(c) $(0.8,1,1,0.8,0,0.3,0.3,0.3,0)$,
and (d) $\left(\frac{1}{\sqrt{2}},1,1,\frac{1}{\sqrt{2}},0,0,0,0,0 \right)$.
Red arrows point to the flat bands, and the blue circle represents the quadruple band touching.
The coordinates of the high-symmetry points in the first Brillouin zone are
$\Gamma = (0,0,0)$, $W= \left(\pi, 0, 2\pi \right)$, $L = \left( \pi, \pi, \pi \right)$, 
$X = \left(0,0, 2\pi \right)$ and $K = \left(\frac{3\pi}{2}, 0, \frac{3\pi}{2}  \right)$.}
  \label{fig:2}
 \end{center}
 \vspace{-10pt}
\end{figure*}

It is also interesting to find that the triple band touching,
where the flat band penetrates the band touching point of dispersive bands,
occurs at $\Gamma$ point in some cases
[e.g., $\varepsilon = 0$ in Fig.~\ref{fig:1}(d)].
In what follows, we elucidate the condition for the triple band touching, 
by explicitly derive the eigenenergies at $\Gamma$ point. 

Before proceeding further, 
we remark that any of the flat bands touches 
the dispersive band at $\Gamma$ point 
regardless of the parameters. 
This is because of the rank reduction of 
$\Psi_{\bm{k}}^\dagger$, which we have mentioned in Sec.~\ref{sec:formulation_M}.
Therefore, from the above derivation of the flat-band energies, 
we have already obtained $2q$ eigenenergies out of $3q + 2$ at $\Gamma$ point,
thus we need to derive the remaining $q+2$ eigenenergies. 

For the derivation of the eigenenergies at $\Gamma$ point, 
we first point out that the remaining eigenstates have three-fold rotational symmetries centered at A site and B site.
Therefore, the wave function satisfies
\begin{eqnarray}
\varphi_{m} =  \varphi_{m+q} = \varphi_{m+2q}, \label{eq:wavecond}
\end{eqnarray}
for $m =1,\cdots, q$.
Substituting (\ref{eq:wavecond}) into the Schr\"{o}dinger equation, we find that it is reduced to 
the eigenvalue equation of the following $(q + 2) \times (q+2)$ matrix:
\begin{eqnarray}
\mathcal{X} = \begin{pmatrix}
U_{\rm A} & 3t_1 & &&& \\
t_1 & U_1 & t_2 &&& \\
 & t_2& U_2 & \ddots & \\
& &  \ddots &\ddots & \\
& & & & U_{q} & t_{q+1} \\
&&&& 3t_{q+1} & U_{\rm B} \\
\end{pmatrix}.
\end{eqnarray}
Clearly, $\mathcal{X}$ is a non-Hermitian matrix,
since $(1,2)$ and $(2,1)$ components are different and so are $(q-1,q)$ and $(q,q-1)$ components.
Nevertheless, all the eigenvalues of $\mathcal{X}$ are real, 
since there exists a similarity transformation such that 
$\mathcal{X}$ is transformed into the Hermitian matrix: 
\begin{eqnarray}
P^{-1}\mathcal{X} P
= \tilde{\mathcal{H}},
\end{eqnarray}
with $P = \mathrm{diag} \left(\sqrt{3}, 1, \cdots, 1, \sqrt{3} \right)$ and 
\begin{eqnarray}
\tilde{\mathcal{H}} = 
\begin{pmatrix}
U_{\rm A} & \sqrt{3} t_1 & &&& \\
\sqrt{3} t_1 & U_1 & t_2 &&& \\
 & t_2& U_2 & \ddots & \\
& &  \ddots &\ddots & \\
& & & & U_{q} & \sqrt{3} t_{q+1} \\
&&&& \sqrt{3} t_{q+1} & U_{\rm B} \\
\end{pmatrix}.
\nonumber \\
\end{eqnarray}
Then, denoting the eigenvalues of 
$\tilde{\mathcal{H}}$ as 
$\varepsilon^{\rm Disp.}_{n^\prime}(t_1,t_2,t_3,\cdots,t_{q}, t_{q+1};U_{\rm  A}, U_1,U_2, \cdots U_{q} ,U_{\rm B})$
with $n^\prime=1,\cdots,q+2$,
we can write down the condition for the triple band touching as
\begin{eqnarray}
&\varepsilon_{\mathrm{linkage},n} (t_2,t_3,\cdots,t_{q}, U_1,U_2, \cdots U_{q})  \nonumber \\
=& \varepsilon^{\rm Disp.}_{n^\prime}(t_1,t_2,t_3,\cdots,t_{q}, t_{q+1}, U_{\rm  A}, U_1,U_2, \cdots U_{q} ,U_{\rm B}), \nonumber \\
\end{eqnarray}
for some $n=1,\cdots ,q$ and $n^\prime = 1, \cdots q+2$.

\begin{figure*}[t]
\begin{center}
\includegraphics[clip,width = 0.7\linewidth]{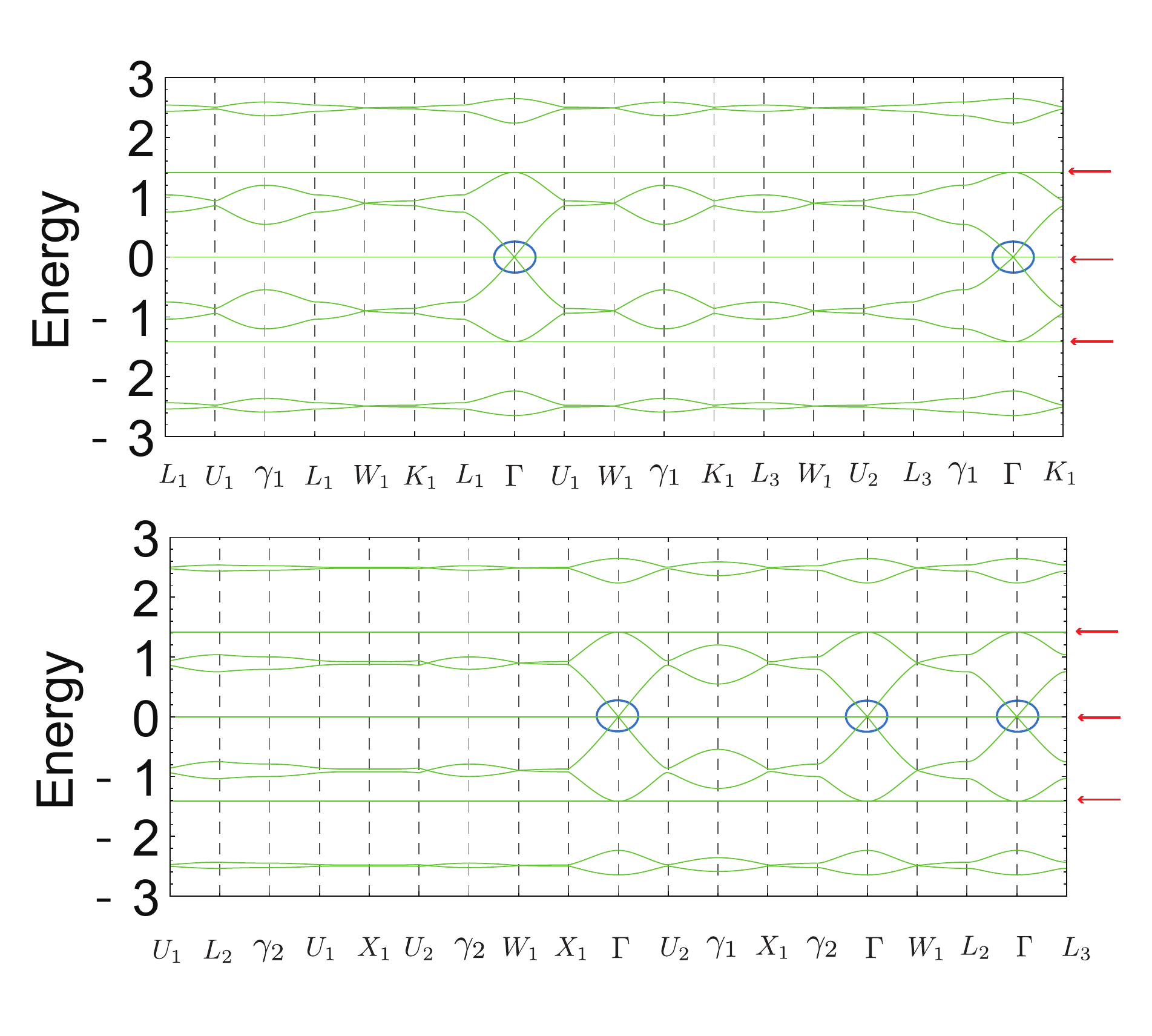}
\vspace{-10pt}
\caption{
The band structure of the decorated four-dimensional diamond lattice with $q=3$.
The parameters are set as 
$(t_1,t_2,t_3,t_4,U_{\rm A},U_1,U_2,U_3,U_{\rm B}) = (1,1,1,1,0,0,0,0,0)$.
Upper and lower panels are for different high-symmetry lines. 
Red arrows point to the flat bands, and blue circles represent the quintuple band touchings.}
  \label{fig:4dd}
 \end{center}
 \vspace{-10pt}
\end{figure*}
For the special case when both $\mathcal{H}_{\mathrm{linkage}}$ 
and $\tilde{\mathcal{H}}$ are chiral symmetric
(i.e., $U_{\rm A} = U_{\rm B} = U_{1} = \cdots =U_{q} =  0$), and $q$ is an odd number, 
both $\mathcal{H}_{\mathrm{linkage}}$ 
and $\tilde{\mathcal{H}}$ have a zero eigenvalue, thus 
the triple band touching at $\varepsilon = 0$ is guaranteed.
Indeed, Fig.~\ref{fig:1}(d) is an example of such a case. 
As a further interesting case, we show an example where 
all of the flat bands exhibit the triple band touching in Fig.~\ref{fig:1}(f).
In fact, such a set of parameters can be found by using the wisdom of the Dynkin diagrams; see Appendix~\ref{app:dynkin} for details.

Before closing this section, we present two additional examples beyond the model discussed so far.
The first one is the case where the decoration sites are not aligned in 
a chain, as depicted in Fig.~\ref{fig:dhsq}(a).
Even in this case, the flat-band energies are obtained by solving the eigenvalue problem of the ``molecule"
formed by the decorating sites [Fig.~\ref{fig:dhsq}(b)].
Indeed, we find four flat bands in Fig.~\ref{fig:dhsq}(c), whose energies are equal to those for Fig.~\ref{fig:dhsq}(b).

The second example is the Chern insulator on the decorated honeycomb lattice. 
As we have seen, the flat band wave functions have vanishing amplitudes on A and B. 
Therefore, if one modifies the model such that the additional term acts only on A and B, 
the model still hosts the exact flat band.
Keeping this in mind, we add the complex hopping among the vertices of the honeycomb lattice
to $\mathcal{H}^{\rm DH}_{\bm{k}}$, 
to make the dispersive bands topological [see Fig.\ref{fig:Chern}~(a) for the schematic figure].
Specifically, the additional term, $\mathcal{H}_{\bm{k}}^\prime$, has the same form as 
the Haldane model~\cite{Haldane1988}:
\begin{eqnarray}
\mathcal{H}^\prime_{\bm{k}} = 2 \lambda M_{\bm{k}} \mathrm{diag}(1, 0,\cdots,0, -1)
\end{eqnarray}
with $M_{\bm{k}} = \sin \bm{k} \cdot \bm{a}_1^{\mathrm{DH}} -  
\sin  \bm{k} \cdot \bm{a}_2^{\mathrm{DH}}
- \sin \bm{k} \cdot \left(\bm{a}_1^{\mathrm{DH}} - \bm{a}_2^{\mathrm{DH}} \right)$.
The band structure for a representative set of parameters with $q=3$ is shown in Fig.~\ref{fig:Chern}(b).
We compute the Chern number numerically by using the method of Ref.~\onlinecite{Fukui2005}.
Clearly, the exact flat bands survive and some of the dispersive bands acquire the non-trivial Chern numbers.
Further, some of the flat bands have quadratic band touching with topologically-nontrivial dispersive bands.
Similar band structure was seen in the kagome-lattice model discussed in Ref.~\onlinecite{Mizoguchi2020}.

\subsection{Three and four dimensions \label{sec:diamond}}
The same method is applicable to the case of three- and four-dimensional decorated honeycomb lattice
with $q$ sites on each edge. In such models, the flat-band energies are given by the eigenenergies of 
$\mathcal{H}_{\rm linkage}$, regardless of the dimensionality.

Figures~\ref{fig:2} and \ref{fig:4dd} show the resulting band structures for 
three and four dimensions,
respectively, with $q=3$. 
For the coordinates of the high-symmetry points in the four-dimensional Brillouin zone,
we follow Ref.~\cite{Kato2017}; see Appendix~\ref{sec:4DBZ}. 
We note that the degeneracy of each flat band is two (three) for $D=3$ ($D=4$).
Correspondingly, the band touchings 
at $\Gamma$ point denoted by the blue circles in Figs.~\ref{fig:2} and \ref{fig:4dd}
have $D+1$-fold degeneracy for the $D$-dimensional system.

\section{$D$-dimensional decorated pyrochlore lattices \label{sec:decorated_py}}
\begin{figure*}[tb]
\begin{center}
\includegraphics[clip,width = 0.95\linewidth]{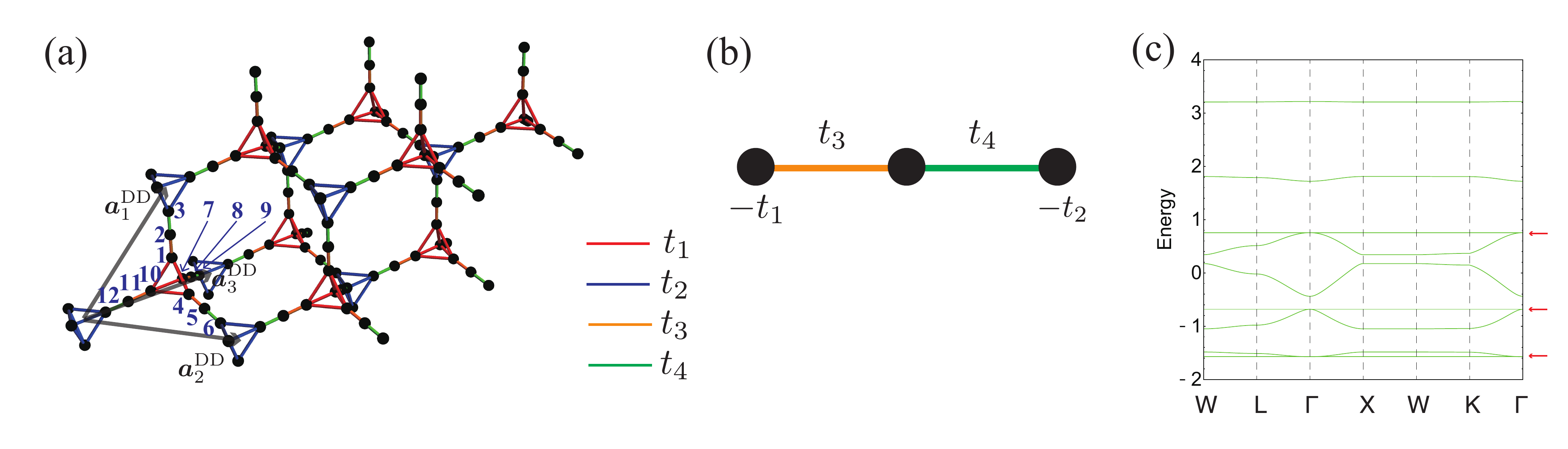}
\vspace{-10pt}
\caption{(a) A decorated pyrochlore lattice with one decorated site between neighboring tetrahedra. 
(b) Schematic figure of the Hamiltonian of the chain-like molecule corresponding to $\mathcal{H}_{\mathrm{linkage}}^{\rm DP}$.
(c) The band structure for $(t_1,t_2,t_3,t_4) = (1,0.5, 0.8,0.7)$. Red arrows point to the flat bands.}
  \label{fig:py}
 \end{center}
 \vspace{-10pt}
\end{figure*}
In this section, we discuss yet another series of multiple flat-band systems, namely, 
$D$-dimensional decorated pyrochlore lattices.
For concreteness, we consider the three-dimensional decorated pyrochlore model 
with one decorating site between neighboring tetrahedra
[Fig.~\ref{fig:py}(a)].
Extension to generic dimensions and generic forms of decoration is straightforward. 
(For instance, the result for the two-dimensional analog is presented in the prior work~\cite{Mizoguchi2019_3}.)
We note that this type of lattice structure, both in two and three dimensions, has various material realizations,
mainly in organic systems~\cite{Fujii2018,Fujii2019,Mizoguchi2019_3,You2019,Sheng2011_T,Janani2014,Peng2021}.

We consider the lattice of Fig.~\ref{fig:py}(a). 
Three lattice vectors are in common with the decorated diamond lattice. 
The Hamiltonian is the $12 \times 12$ matrix given as
\begin{eqnarray}
\mathcal{H}^{\rm DP}_{\bm{k}} = 
\begin{pmatrix}
\tilde{\mathcal{H}}^{\rm DP}_{\mathrm{linkage}} & V_{\bm{k},(1,2)} &  V_{\bm{k},(1,3)}&  V_{\bm{k},(1,4)} \\ 
V_{\bm{k},(2,1)}& \tilde{\mathcal{H}}^{\rm DP}_{\mathrm{linkage}}  &  V_{\bm{k},(2,3)} & V_{\bm{k},(2,4)} \\
V_{\bm{k},(3,1)}&  V_{\bm{k},(3,2)} & \tilde{\mathcal{H}}^{\rm DP}_{\mathrm{linkage}}& V_{\bm{k},(3,4)} \\
V_{\bm{k},(4,1)} & V_{\bm{k},(4,2)} & V_{\bm{k},(4,3)} & \tilde{\mathcal{H}}^{\rm DP}_{\mathrm{linkage}}\\
\end{pmatrix}, \label{eq:ham_gen_py}
\end{eqnarray}
where 
\begin{eqnarray}
\tilde{\mathcal{H}}^{\rm DP}_{\mathrm{linkage}}  
= \begin{pmatrix}
0 & t_3 & 0 \\
t_3 & 0 & t_4 \\
0 & t_4 & 0 \\
\end{pmatrix}, \label{eq:DP_linkage_1}
\end{eqnarray}
and 
\begin{eqnarray}
V_{\bm{k},(i,j)} = 
\begin{pmatrix}
t_1 & 0 & 0 \\
0 & 0 & 0 \\
0& 0 &t_2 e^{-i \bm{k}\cdot (\bm{a}^{\mathrm{DD}}_{i}-\bm{a}^{\mathrm{DD}}_j)}
\end{pmatrix}
\end{eqnarray}
with $\bm{a}_4^{\mathrm{DD}} = (0,0,0)$.

To obtain the flat band solution, we again give the intertwiner explicitly. 
In the present model, we have 
\begin{eqnarray}
\mathcal{H}^{\rm DP}_{\bm{k}}  C_{\bm{k}} 
= C_{\bm{k}} \mathcal{H}_{\mathrm{linkage}}^{\rm DP}, \label{eq:intertwiner_DP}
\end{eqnarray}
where
\begin{eqnarray}
C_{\bm{k}} = \begin{pmatrix}
[\bm{\lambda}_{\mathrm{linker},\bm{k}}]_1 I_3 \\
[\bm{\lambda}_{\mathrm{linker},\bm{k}}]_2 I_3 \\
[\bm{\lambda}_{\mathrm{linker},\bm{k}}]_3 I_3 \\
[\bm{\lambda}_{\mathrm{linker},\bm{k}}]_4 I_3 \\
\end{pmatrix}, \label{eq:dpy_intertwiner}
\end{eqnarray}
and 
\begin{eqnarray}
\mathcal{H}_{\mathrm{linkage}}^{\rm DP}&=& 
\tilde{\mathcal{H}}_{\mathrm{linkage}}^{\rm DP}
+ \begin{pmatrix}
-t_1 & 0 & 0 \\
0 & 0 & 0 \\
0 & 0 & -t_2 \\
\end{pmatrix}
\nonumber \\
&=&
\begin{pmatrix}
-t_1 & t_3 & 0 \\
t_3  & 0 & t_4 \\
0 & t_4 & -t_2 \\
\end{pmatrix}, \label{eq:DP_linkage_2}
\end{eqnarray}
where $\bm{\lambda}_{\mathrm{linker},\bm{k}}$ is the same as that for the decorated diamond model. 
Note that the left-hand side of Eq.~(\ref{eq:intertwiner_DP}) becomes
\begin{widetext}
\begin{eqnarray}
\mathcal{H}^{\rm DP}_{\bm{k}}  C_{\bm{k}} &=&
\begin{pmatrix}
\tilde{\mathcal{H}}^{\rm DP}_{\mathrm{linkage}} & V_{\bm{k},(1,2)} &  V_{\bm{k},(1,3)}&  V_{\bm{k},(1,4)} \\ 
V_{\bm{k},(2,1)}& \tilde{\mathcal{H}}^{\rm DP}_{\mathrm{linkage}}  &  V_{\bm{k},(2,3)} & V_{\bm{k},(2,4)} \\
V_{\bm{k},(3,1)}&  V_{\bm{k},(3,2)} & \tilde{\mathcal{H}}^{\rm DP}_{\mathrm{linkage}}& V_{\bm{k},(3,4)} \\
V_{\bm{k},(4,1)} & V_{\bm{k},(4,2)} & V_{\bm{k},(4,3)} & \tilde{\mathcal{H}}^{\rm DP}_{\mathrm{linkage}}\\
\end{pmatrix}
\begin{pmatrix}
[\bm{\lambda}_{\mathrm{linker},\bm{k}}]_1 I_3 \\
[\bm{\lambda}_{\mathrm{linker},\bm{k}}]_2 I_3 \\
[\bm{\lambda}_{\mathrm{linker},\bm{k}}]_3 I_3 \\
[\bm{\lambda}_{\mathrm{linker},\bm{k}}]_4 I_3 \\
\end{pmatrix} \nonumber \\
&=& \begin{pmatrix}
[\bm{\lambda}_{\mathrm{linker},\bm{k}}]_1 \tilde{\mathcal{H}}^{\rm DP}_{\mathrm{linkage}} 
+  [\bm{\lambda}_{\mathrm{linker},\bm{k}}]_2 V_{\bm{k},(1,2)} + [\bm{\lambda}_{\mathrm{linker},\bm{k}}]_3 V_{\bm{k},(1,3)} + 
[\bm{\lambda}_{\mathrm{linker},\bm{k}}]_4 V_{\bm{k},(1,4)} \\
[\bm{\lambda}_{\mathrm{linker},\bm{k}}]_2 \tilde{\mathcal{H}}^{\rm DP}_{\mathrm{linkage}} 
+  [\bm{\lambda}_{\mathrm{linker},\bm{k}}]_1 V_{\bm{k},(2,1)} + [\bm{\lambda}_{\mathrm{linker},\bm{k}}]_3 V_{\bm{k},(2,3)} + 
[\bm{\lambda}_{\mathrm{linker},\bm{k}}]_4 V_{\bm{k},(2,4)} \\
[\bm{\lambda}_{\mathrm{linker},\bm{k}}]_3 \tilde{\mathcal{H}}^{\rm DP}_{\mathrm{linkage}} 
+  [\bm{\lambda}_{\mathrm{linker},\bm{k}}]_1 V_{\bm{k},(3,1)} + [\bm{\lambda}_{\mathrm{linker},\bm{k}}]_2 V_{\bm{k},(3,2)} + 
[\bm{\lambda}_{\mathrm{linker},\bm{k}}]_4 V_{\bm{k},(3,4)} \\
[\bm{\lambda}_{\mathrm{linker},\bm{k}}]_4 \tilde{\mathcal{H}}^{\rm DP}_{\mathrm{linkage}} 
+  [\bm{\lambda}_{\mathrm{linker},\bm{k}}]_1 V_{\bm{k},(4,1)} + [\bm{\lambda}_{\mathrm{linker},\bm{k}}]_2 V_{\bm{k},(4,2)} + 
[\bm{\lambda}_{\mathrm{linker},\bm{k}}]_3 V_{\bm{k},(4,3)} \\
\end{pmatrix}. \label{eq:henkei}
\end{eqnarray} 
The $j$-th column of the second line of Eq.~(\ref{eq:henkei}) is 
\begin{eqnarray}
&& [\bm{\lambda}_{\mathrm{linker},\bm{k}}]_j \tilde{\mathcal{H}}^{\rm DP}_{\mathrm{linkage}} 
+\sum_{j^\prime \neq j} [\bm{\lambda}_{\mathrm{linker},\bm{k}}]_{j^\prime} 
\begin{pmatrix}
t_1 & 0 & 0 \\
0 & 0 & 0 \\
0& 0 &t_2 e^{-i \bm{k}\cdot (\bm{a}^{\mathrm{DD}}_{j}-\bm{a}^{\mathrm{DD}}_{j^\prime})}
\end{pmatrix} \nonumber \\
&=& [\bm{\lambda}_{\mathrm{linker},\bm{k}}]_j 
\left[ \tilde{\mathcal{H}}^{\rm DP}_{\mathrm{linkage}}
+
\begin{pmatrix}
- t_1 & 0 & 0 \\
0 & 0 & 0 \\
0& 0 & - t_2 \\
\end{pmatrix}
\right] = [\bm{\lambda}_{\mathrm{linker},\bm{k}}]_j \mathcal{H}_{\mathrm{linkage}}^{\rm DP}, \label{eq:henkei2}
\end{eqnarray}
\end{widetext}
which is equal to the $j$-th component of the right-hand side of Eq.~(\ref{eq:dpy_intertwiner}).
The second line of Eq.~(\ref{eq:henkei2}) can be obtained by using Eq.~(\ref{eq:def_lambda}).
Having Eq.~(\ref{eq:intertwiner_DP}) at hand,
we again see that the flat-band eigenenergies are equal to those of $\mathcal{H}_{\mathrm{linkage}}^{\rm DP}$,
and that the wave function of $n$th flat band is given as
\begin{eqnarray}
\varphi^{\rm DP}_{3(j-1) + m ,\bm{k},n} = 
\frac{1}{\mathcal{N}_{\bm{k}}} 
[\bm{\lambda}_{\mathrm{linker},\bm{k}}]_j  [\bm{\phi}^{\rm DP}_{\mathrm{linkage},n}]_m,
\label{eq:wave_function_DP}
\end{eqnarray}
($j=1,2,3,4$, $m=1,2,3$), 
where $\bm{\phi}^{\rm DP}_{\mathrm{linkage},n}$ is the eigenvector of $\mathcal{H}_{\mathrm{linkage}}^{\rm DP}$
corresponding to the $n$th eigenvalue.
The corresponding molecule for $\mathcal{H}_{\mathrm{linkage}}^{\rm DP}$ is depicted in Fig.~\ref{fig:py}(b).
Comparing $\tilde{\mathcal{H}}^{\rm DP}_{\mathrm{linkage}}$ with $\mathcal{H}_{\mathrm{linkage}}^{\rm DP}$,
one finds that the on-site potentials, $-t_1$ and $-t_2$, are added at the end sites.  

The band structure for a certain set of parameters is shown in Fig.~\ref{fig:py}(c).
We obtain three flat bands, each of which is doubly degenerate.
As we have discussed, their energies are equal to the eigenvalues of $\mathcal{H}_{\mathrm{linkage}}^{\rm DP}$.

\section{Summary and discussions \label{sec:summary}}
We have presented the method to determine the flat-band energies and wave functions analytically
in the decorated diamond lattices in arbitrary dimensions. 
The key idea is to divide the Hamiltonian into the linker part and the linkage part.
Namely, by using the intertwiner [Eq.~(\ref{eq:intertwiner_fb})] which is composed of the wave functions at the linker, 
we can reduce the eigenvalue problem of 
$\bm{k}$-dependent $[(D+1)q + 2] \times [(D+1)q + 2]$ matrix ($\mathcal{H}_{\bm{k}}$)
to the $\bm{k}$-independent $q \times q$ linkage Hamiltonian ($\mathcal{H}_{\rm linkage}$).
Further, we also find that the flat-band wave function of the $D$-dimensional decorated diamond lattice
is given by the product of the linkage wave function and the flat-band wave function for the $D$-dimensional pyrochlore lattice.

We show the examples of the decorated honeycomb lattice in two dimensions,
the decorated diamond lattice in three dimensions, and the decorated four-dimensional diamond lattice,
where each NN bond is decorated by the chain-like structure.
The condition for the multiple band touching at $\Gamma$ point is also addressed. 
Further, the same method is applicable to the $D$-dimensional decorated pyrochlore lattices.
There, the tetrahedral parts of the original Hamiltonian turn into the on-site potential 
at the edges of the linkage Hamiltonian.

As mentioned in Sec.~\ref{sec:intro}, several extensions of our method are possible, as listed below.
\begin{itemize} 
\item[(i)] We assume that each linkage is connected to a linker through one of the sites. 
However, this method can be used even when each linkage is connected to a linker with more than two sites 
[for an example, see Fig.~\ref{fig:ex}(a)].
This is because the relation Eq.~(\ref{eq:intertwiner_fb}) for the intertwiner of Eq.~(\ref{eq:intertwiner}) 
holds even in this case.
\item[(ii)] We assume that all the hopping integrals in each linker are the same.
This condition can be relaxed, i.e., the hopping integrals in each linker can be different [for an example, see Fig.~\ref{fig:ex}(c)].
\item[(iii)] We assume that all the linkages have the same structure [Eq.~(\ref{eq:linkage_same})].
However, our construction of the flat bands works even when linkages have different structures, 
as long as the linkages have common eigenenergies~\cite{Katsura2015}.
For instance, the numbers of sites consisting of the linkages can be different from each other [see Fig.~\ref{fig:ex}(e)].
\item[(iv)] Finally, the lattices structures are not limited to the decorated diamond lattices.
In fact, the method works in, e.g., the decorated square lattices (i.e., the generalized Lieb lattices)~\cite{Katsura2015}.
In this regard, the Lieb-lattice-based materials are also in the scope of application of this method~\cite{Cui2020,Mao2020,Liu2021}.
\end{itemize}
Although the comprehensive descriptions about the generalizations are beyond the scope of this paper,
we show some of the results of the generalized models in Appendix~\ref{app:extensions}.

To conclude, there are a number 
of materials with decorated honeycomb, diamond and pyrochlore lattice structures, especially for organic materials.
We hope that our method to determine flat-band energies and wave functions
is useful for band structure analysis and material design. 

\textit{Note added.}--- 
Recently, we became aware of the related works~\cite{Qi2020,Boudjada2020} 
where the flat bands of the decorated honeycomb model are discussed.

 \acknowledgments
This work is supported by JSPS KAKENHI, Grants No.~JP17H06138 (T. M. and Y. H.) 
and No.~JP20K14371 (T. M.).
H. K. was supported in part by JSPS Grant-in-Aid for Scientific Research on Innovative Areas 
No.~JP20H04630, JSPS KAKENHI Grant No. JP18K03445, and the Inamori Foundation. 

\appendix
\section{Examples of the extended models \label{app:extensions}}
\begin{figure}[tb]
\begin{center}
\includegraphics[clip,width = 0.95\linewidth]{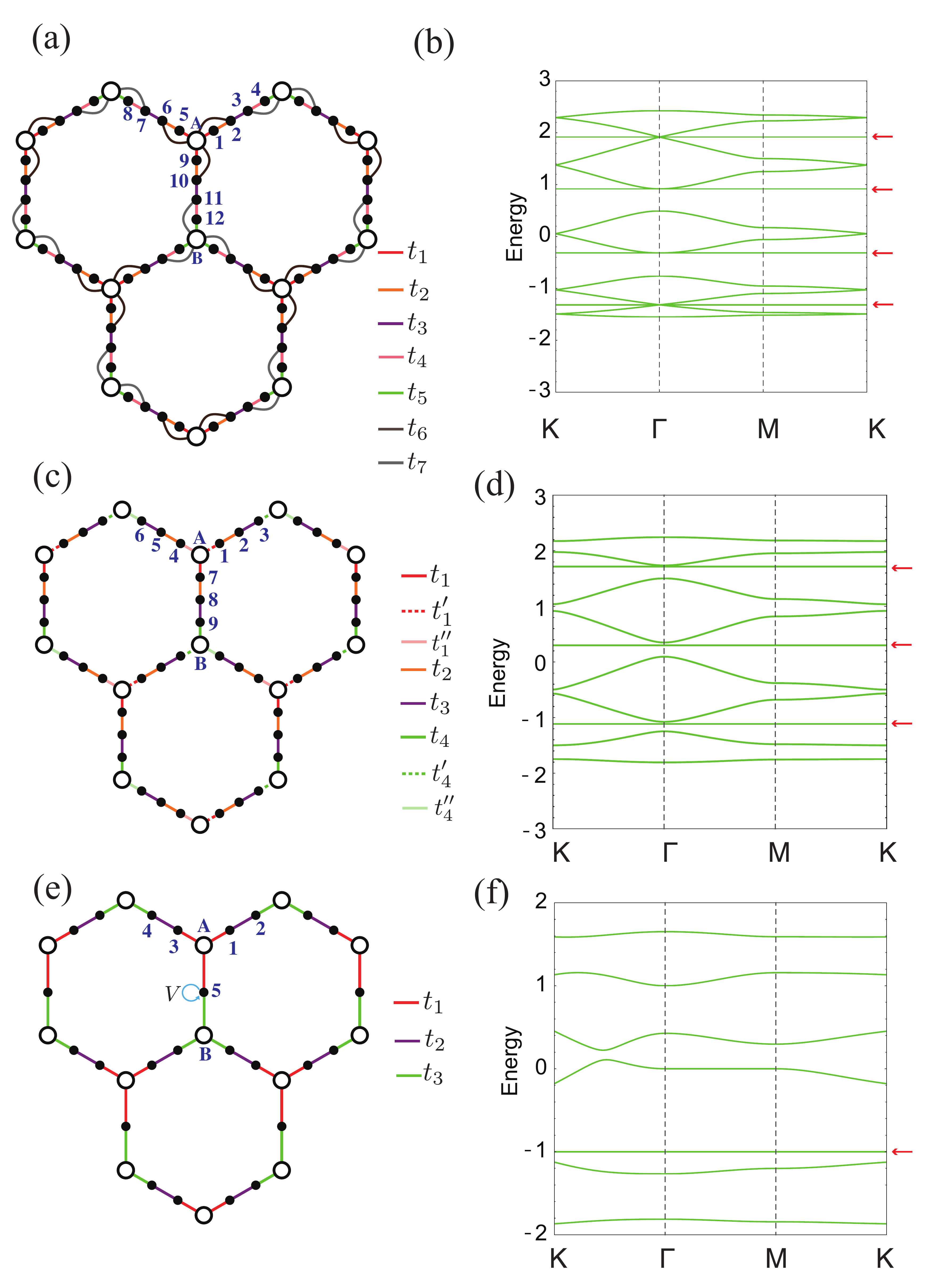}
\vspace{-10pt}
\caption{
The examples of the generalizations of the flat-band models on decorated honeycomb lattices.
(a) Model where two sites on each linkage are connected to a linker.
(b) Band structure for the model of (a). 
The parameters are set as $(t_1,t_2,t_3,t_4,t_5,t_6,t_7,U_{\rm A},U_1,U_2,U_3,U_4,U_{\rm B}) = (0.8,1,1,1,0.8,0.2,0.2,0,0.3,0.3,0.3,0.3,0)$ 
(The definitions of the above parameters follow those in Sec.~\ref{sec:honeycomb}). 
(c) Model where the hoppings integrals on each linker are different. 
(d) Band structure for the model of (c). 
The parameters are set as $(t_1,t_1^{\prime},t_1^{\prime \prime}, 
t_2,t_3,t_4,t_4^\prime,t_4^{\prime \prime},U_{\rm A},U_1,U_2,U_3,U_{\rm B}) = (0.8,0.7,0.6,1,1,0.8,0.9,1,0,0.3,0.3,0.3,0)$. 
(e) Model where the linkages have the different structures from each other. 
(f) Band structure for the model of (e). 
The hopping parameters are set as $(t_1,t_2,t_3) = (0.8,1,0.4)$.
The on-site potential is introduced only at the sublattice 5 with the energy $V = -1$.
Red arrows point to the flat bands.}
  \label{fig:ex}
 \end{center}
 \vspace{-10pt}
\end{figure}
In this appendix, we show two examples where the assumptions described in Sec.~\ref{sec:formulation} are relaxed.
Here we focus on the case of $D=2$.

The first model is depicted in Fig.~\ref{fig:ex}(a) (we set $q=4$), 
where two sites on each linkage are connected to a linker.
Specifically, the second-neighbor hoppings 
$t_6$ and $t_7$ are included in addition to the NN hoppings. 
The band structure for a representative set of parameters is shown in Fig.~\ref{fig:ex}(b).
We see that there exist four exact flat bands.
In fact, the flat-band energies and eigenvectors are given in exactly the same forms as described in the main text
since Eq.~(\ref{eq:intertwiner_fb}) for the intertwiner of Eq.~(\ref{eq:intertwiner}) holds even in this case.
Therefore, the flat bands are not affected by the inclusion of the second-neighbor hoppings of this kind.

The second model is depicted in Fig.~\ref{fig:ex}(c) (we set $q=3$),
where the hopping integrals in each linker are different from each other.
For instance, the linker including sublattice A contains three different hoppings, $t_1$, $t_1^\prime$, and $t_1^{\prime \prime}$.
The band structure for a representative set of parameters is shown in Fig.~\ref{fig:ex}(d).
We see three exact flat bands.
In fact, the flat bands can be obtained by replacing $\bm{\lambda}_{\mathrm{linker},\bm{k}}$ in the intertwiner of 
Eq.~(\ref{eq:intertwiner})
with $\tilde{\bm{\lambda}}_{\mathrm{linker},\bm{k}}$,
which satisfies
\begin{eqnarray}
\begin{pmatrix}
t_1^\prime &  t_1^{\prime \prime} & t_1 \\
t_4^\prime e^{- i\bm{k} \cdot \bm{a}_1^{\rm DH}}  & t_4^{\prime \prime} e^{-i\bm{k} \cdot \bm{a}_2^{\rm DH}} & t_4 \\
\end{pmatrix}
 \tilde{\bm{\lambda}}_{\mathrm{linker},\bm{k}} = 
\begin{pmatrix}
0 \\
0\\
\end{pmatrix}.
\end{eqnarray} 

The third model is depicted in Fig.~\ref{fig:ex}(e), where the linkages are not the same.
Specifically, two of three linkages around A have $q = 2$ whereas the other has $q = 1$.
Only for the linkage of $q = 1$, we introduce the on-site potential $V$ so that all three linkages have a common eigenenergy.
The band structure for a representative set of parameters is shown in Fig.~\ref{fig:ex}(f).
We see that there is an exact flat band, whose energy is the same as the common eigenenergy of the linkages.

\section{Specific cases with triple band touching \label{app:dynkin}}
\begin{figure}[tb]
\begin{center}
\includegraphics[clip,width = 0.95\linewidth]{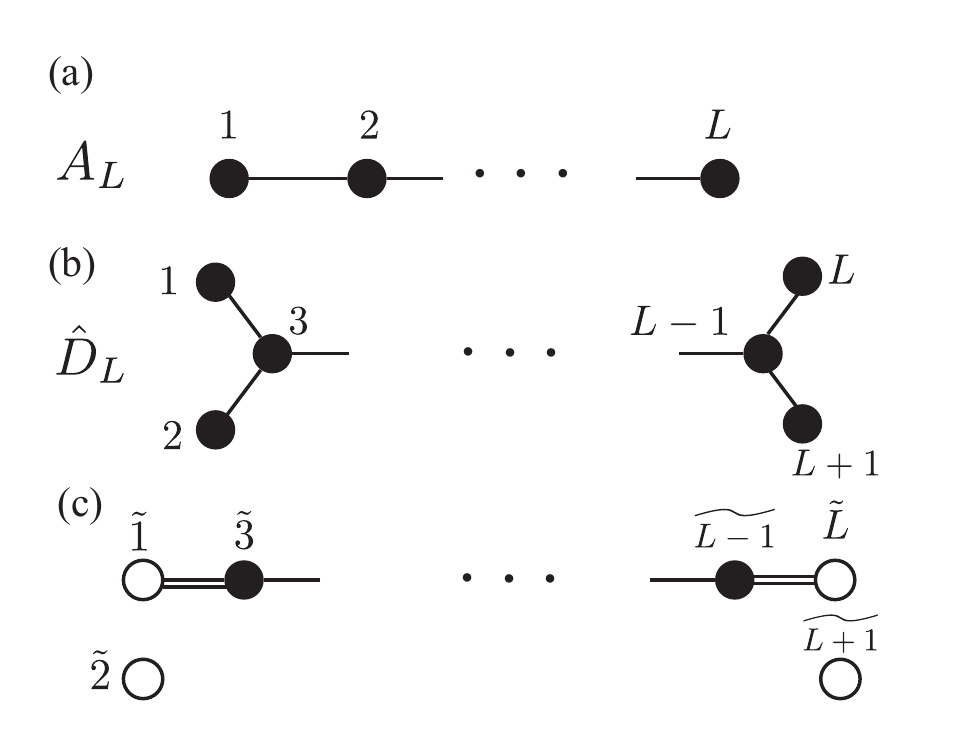}
\vspace{-10pt}
\caption{The Dynkin diagram of (a) $A_L$ and the extended Dynkin diagram of (b) ${\hat D}_L$.
(c) Schematic figure of the chain plus two isolated sites equivalent to $\hat{D}_L$ under the change of the basis. 
The single lines denote the bonds with the hopping being unity, while the double lines denote the bonds with the hopping being $\sqrt{2}$. }
  \label{fig:dyn}
 \end{center}
 \vspace{-10pt}
\end{figure}
In this appendix, we elucidate that the condition for the triple band touching at $\Gamma$ 
point in the decorated honeycomb model can be found exactly for the special case. 
Specifically, we restrict ourselves to the case where 
$U_{\rm A} = U_{\rm B} = U_{1} = \cdots = U_{q} =0$,
$t_2 = \cdots t_{q} = 1$, and $t_{1} = t_{q+1} = \tilde{t}$. 
The aim here is to determine 
$\tilde{t}$ such that all of the $q$ flat bands are involved in triple band touching at $\Gamma$ point,
as shown in Fig.~\ref{fig:1}(f).

To this aim, we employ the wisdom of the eigenvalues of the adjacency matrices of the Dynkin diagrams (or A-D-E lattices). 
Specifically, for the present purpose, we consider the $A$ type [Fig.~\ref{fig:dyn}(a)], which is nothing but the open chain, 
and the $\hat{D}$ type [Fig.~\ref{fig:dyn}(b)], which has double branches at both ends. 
It is known~\cite{Brouwer,Henkel} that the eigenvalues of the adjacency matrix of
$A_L$ are given as 
\begin{eqnarray}
\varepsilon^{A_L}  = 2 \cos \frac{j \pi}{L+1}\hspace{1mm}(j=1, \cdots L), \label{eq:eigen_AL}
\end{eqnarray}
while those of $\hat{D}_L$ are given as 
\begin{eqnarray}
\varepsilon^{\hat{D}_{L}} = 0, 2 \cos \frac{j \pi}{L-2} \hspace{1mm}(j=0, \cdots L-2).\label{eq:eigen_DL}
\end{eqnarray}
From Eqs.~(\ref{eq:eigen_AL}) and (\ref{eq:eigen_DL}), 
we see that all of the eigenvalues for 
$A_L$ are included in the set of the eigenvalues of $\hat{D}_{L+3}$.
We note that this fact can also be derived by explicitly giving the intertwiner between the adjacency matrices for these graphs.
Namely, the following relation holds:
\begin{eqnarray}
H_{\hat{D}_{L+3}} C_{L} = C_{L} H_{A_L},
\end{eqnarray}
with
\begin{eqnarray}
\left(C_{L}\right)_{ij} &=& \delta_{i,1} \delta_{j,1} + \delta_{i,j+1} - \delta_{i,j+3} - \delta_{i,L+4} \delta_{j,L} \nonumber \\
(i &=&1,\cdots L+4,\hspace{1mm} j=1,\cdots L),
\end{eqnarray}
where $H_{\hat{D}_{L+3}}$ and $H_{A_L}$ stand for the adjacency matrices of $\hat{D}_{L+3}$ and $A_L$, respectively.

Further, as for $\hat{D}_{L}$, by changing the basis as 
$\ket{\tilde{1}} = \frac{1}{\sqrt{2}} \left[\ket{1} +\ket{2} \right]$,
$\ket{\tilde{2}} = \frac{1}{\sqrt{2}} \left[\ket{1} -\ket{2} \right]$,
$\ket{\tilde{L}} = \frac{1}{\sqrt{2}} \left[\ket{L} + \ket{L+1} \right]$,
$\ket{\widetilde{L+1}} = \frac{1}{\sqrt{2}} \left[\ket{L} - \ket{L+1} \right]$,
and $\ket{\tilde{\ell} } = \ket{\ell}$ ($\ell = 3, \cdots, L-1$),
where $\ket{\ell}$ denotes the state localized at the $\ell$th site in the original graph,
one can see that the hopping problem on the graph $\hat{D}_{L}$ is equivalent to 
that on the $(L-1)$-site chain where the hoppings on the both of the ends are modulated from $1$ to $\sqrt{2}$ 
[see the double lines in Fig.~\ref{fig:dyn}(c)].

Combining these facts, we find the following: 
All of the eigenenergies of the $q$-site chain with the NN hopping being $1$ 
are included in the set of the eigenenergies of the $q+2$-site chain 
where the hoppings are $\sqrt{2}$ on the both of the ends and 1 otherwise.
Turning to our original problem, 
we find that the multiple triple band touchings can be found by setting $\sqrt{3} \tilde{t} = \sqrt{2}$,
which leads to $\tilde{t} = \sqrt{\frac{2}{3}}$.
This is indeed the parameters employed for Fig.~\ref{fig:1}(f) (for $q=3$).
It is to be stressed that the condition for $\tilde{t}$ obtained here is regardless of $q$.
In fact, for $q=2$, the multiple triple band touchings were found in Ref.~\onlinecite{Barreteau2017} 
for the same parameter choice.
We also note that, 
for $D$-dimensional systems, the multiple band touchings whose degeneracy is $D+1$ can be found for 
$\tilde{t} = \sqrt{\frac{2}{D+1}}$.
An example of $D=3$ is shown in Fig.~\ref{fig:2}(d).

\section{High symmetry points of the first Brillouin zone in the four-dimensional diamond lattice \label{sec:4DBZ}}
The four lattice vectors of the four-dimensional diamond lattice are
\begin{eqnarray}
\bm{a}_1^{\rm 4DD} = \left( \frac{\sqrt{5}}{4},\frac{\sqrt{5}}{4},\frac{\sqrt{5}}{4},\frac{5}{4} \right), 
\end{eqnarray}
\begin{eqnarray}
\bm{a}_2^{\rm 4DD} =  \left( \frac{\sqrt{5}}{4},-\frac{\sqrt{5}}{4},-\frac{\sqrt{5}}{4},\frac{5}{4} \right), 
\end{eqnarray}
\begin{eqnarray}
\bm{a}_3^{\rm 4DD} =  \left( -\frac{\sqrt{5}}{4},-\frac{\sqrt{5}}{4},\frac{\sqrt{5}}{4},\frac{5}{4} \right),
\end{eqnarray}
and
\begin{eqnarray} 
\bm{a}_4^{\rm 4DD} =  \left( -\frac{\sqrt{5}}{4},\frac{\sqrt{5}}{4},-\frac{\sqrt{5}}{4},\frac{5}{4} \right).
\end{eqnarray}

For the coordinates of the high-symmetry points in the first Brillouin zone in the four-dimensional diamond lattice,
we follow Ref.~\onlinecite{Kato2017}:
\begin{eqnarray}
\Gamma = (0,0,0,0), \\
\gamma_1 = \left(0,0,0,-\frac{4\pi}{5} \right), \\
\gamma_2 = \left(\frac{2\pi}{\sqrt{5}},0,0,-\frac{2\pi}{5} \right), \\
L_1 = \left(\frac{4\pi}{5 \sqrt{5}}, -\frac{4\pi}{5 \sqrt{5}}, \frac{4\pi}{5 \sqrt{5}}, - \frac{4\pi}{5} \right),\\
L_2 = \left(\frac{2\pi}{\sqrt{5}}, -\frac{2\pi}{5 \sqrt{5}}, \frac{2\pi}{5 \sqrt{5}}, - \frac{2\pi}{5} \right),\\
L_3 = \left(\frac{4\pi}{5 \sqrt{5}}, \frac{4\pi}{5 \sqrt{5}}, \frac{4\pi}{5 \sqrt{5}}, - \frac{4\pi}{5} \right),\\
W_1 = \left(\frac{8\pi}{5\sqrt{5}},0,\frac{4\pi}{5\sqrt{5}},-\frac{4\pi}{5} \right),\\
K_1 = \left( \frac{6\pi}{5\sqrt{5}},0,\frac{6\pi}{5\sqrt{5}},-\frac{4\pi}{5}\right), \\
X_1 = \left( \frac{8\pi}{5\sqrt{5}},0,0, -\frac{4\pi}{5} \right),\\
U_1 = \left(\frac{8\pi}{5\sqrt{5}}, -\frac{2\pi}{5\sqrt{5}}, \frac{2\pi}{5\sqrt{5}},-\frac{4\pi}{5} \right),\\
U_2 = \left( \frac{8\pi}{5\sqrt{5}}, \frac{2\pi}{5\sqrt{5}}, \frac{2\pi}{5\sqrt{5}},-\frac{4\pi}{5}\right).
\end{eqnarray}


\begin{thebibliography}{1}
\bibitem{remark}
Note that $\Psi_{\bm{k}}^\dagger$ was called the ``molecular orbital" (MO) for the $D$-dimensional pyrochlore model~\cite{Hatsugai2011},
by which the Hamiltonian of the tight-binding model with 
NN hoppings being $t$ can be written as 
$\mathcal{H}^{\rm Py}_{\bm{k}} = t \left[ \Psi_{\bm{k}} \Psi^\dagger_{\bm{k}} - 2 I_{D+1} \right]$, with $I_{D+1}$ being the $(D+1) \times (D+1)$ identity matrix.
\end{thebibliography}
\end{document}